\documentclass{article}
\usepackage{spconf,amsmath,graphicx}
\usepackage{hyperref}

\title{PromptSpeaker: Speaker Generation Based on Text Descriptions}
\name{Yongmao Zhang$^1$, Guanghou Liu$^1$, Yi Lei$^1$, Yunlin Chen$^2$, Hao Yin$^2$, Lei Xie$^1$$^*$, Zhifei Li$^2$}

\address{
  $^1$Audio, Speech and Language Processing Group (ASLP@NPU)\\School of Computer Science,
  Northwestern Polytechnical University, Xi’an, China\\
  $^2$Shanghai Mobvoi Information Technology Co., Ltd}
\begin{document}
%

\maketitle
\begin{abstract}

\renewcommand{\thefootnote}{\fnsymbol{footnote}}
\footnotetext{* Corresponding author.}
Recently, text-guided content generation has received extensive attention. In this work, we explore the possibility of text description-based speaker generation, i.e., using text prompts to control the speaker generation process. Specifically, we propose \textit{PromptSpeaker}, a text-guided speaker generation system. PromptSpeaker consists of a prompt encoder, a zero-shot VITS, and a Glow model, where the prompt encoder predicts a prior distribution based on the text description and samples from this distribution to obtain a semantic representation. The Glow model subsequently converts the semantic representation into a speaker representation, and the zero-shot VITS finally synthesizes the speaker's voice based on the speaker representation. We verify that PromptSpeaker can generate speakers new from the training set by objective metrics, and the synthetic speaker voice has reasonable subjective matching quality with the speaker prompt. \renewcommand{\thefootnote}{\arabic{footnote}}Our audio samples are available on the demo website~\footnote[1]{Demo: \url{https://promptspeaker.github.io/demo/}}.
\end{abstract}
In recent years, speech synthesis has made significant improvements driven by deep learning~\cite{ref_survey}, which encourages the synthetic speech natural as human-like in closed domains. Typical speech synthesis models usually generate speech in the speaker's voice included in the training data. But these models obviously can't satisfy the demand for custom voices in real-world scenarios since it's difficult to obtain speech training data for novel speakers. Moreover, for applications that require many distinct voices, such as virtual characters in games, voice selection for speaker anonymization as well as personalized voice assistants, collecting speech training data of each target speaker always incurs a high cost, which is not a practical solution.



Aiming at synthesizing speech in nonexistent human-sounding voices, \textit{speaker generation}~\cite{ref_speakergeneration} is proposed recently as an effective solution for custom voice generation. Speaker generation has attracted increasing attention in recent years~\cite{ref_createnewvoice,ref_deepgassian}, with the wide applications of multi-speaker speech synthesis. Since speaker identity is typically represented as a high-dimensional vector, extracted by a speaker encoder or speaker ID, the speaker generation task can be simplified to generate a new vector of speaker representation in a multi-speaker TTS system. In light of this idea, TacoSpawn~\cite{ref_speakergeneration} utilizes a parametric prior distribution to represent speaker embeddings for speaker generation. The distribution makes producing a new speaker's voice possible through the sampling process. Although such methods can generate new voices and achieve coarse control from the gender and geographical attributes of the generated speakers, they are limited in precise control from the natural language of speaker's characteristics, such as \textit{age} and fine-grained speaking styles like \textit{croaky} or \textit{breathy}.



For controlling generated speaker's timbre more precisely, the goal of \textit{controllable speaker generation} is to produce desired new voices according to given conditions. In real-world applications, it is natural for users to describe the character of a desired speaker using natural language, e.g., ``\textit{I want a husky voice from a middle-aged man}". Therefore, drawing inspiration from recent advances in language-guided image generation~\cite{ref_dell}, audio generation~\cite{ref_audioldm, ref_makeanaudio} and style-controlled speech synthesis~\cite{ref_styletag, ref_instructtts,ref_promptstyle}, we propose a method that can generate desired speaker's voices based on text descriptions. Since the speaker timbre's description is subjective, it is difficult to extend the data in some ways in the audio generation task, such as constructing new annotated data by concatenating two annotated audio segments~\cite{ref_makeanaudio}. The establishment of cross-modal mappings between semantic representations and speaker representations is also difficult.



\begin{figure*}[ht]
	\centering
	\includegraphics[scale=0.45]{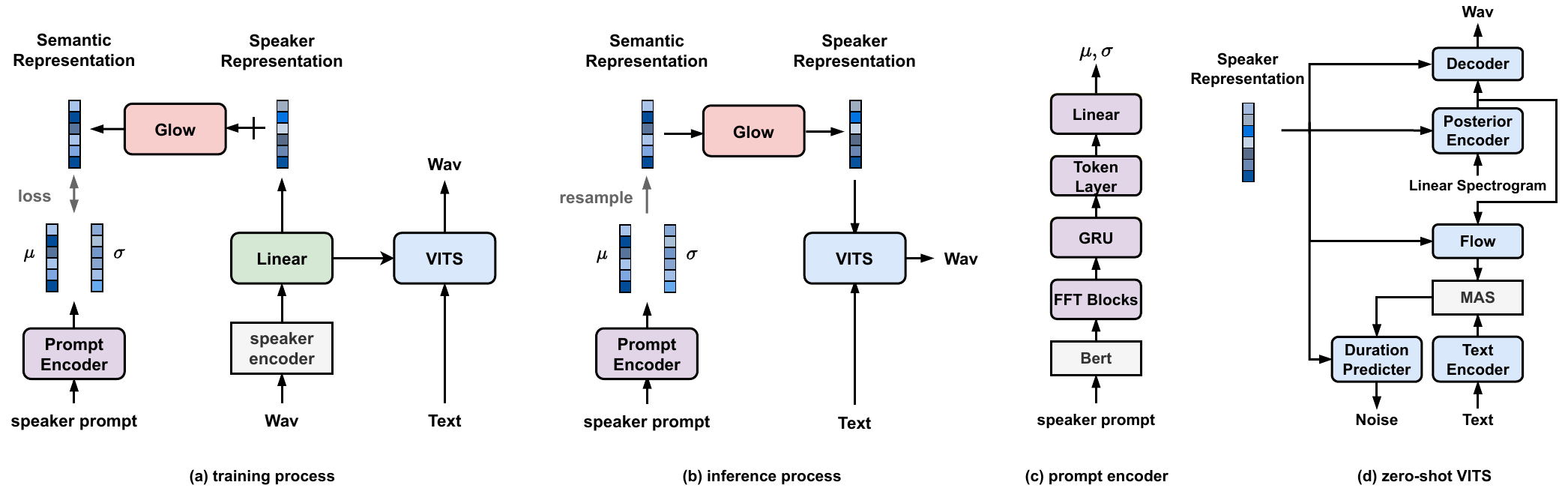}
	\vspace{-2pt}
	\caption{
		Architecture of the PromptSpeaker. 
	}
	\label{model}
\end{figure*}

In this paper, we propose PromptSpeaker, a controllable speaker generation model based on text prompts. PromptSpeaker is composed of a prompt encoder, a Glow model and a zero-shot TTS system. The prompt encoder synthesizes the semantic prior distribution of speaker. The zero-shot TTS system takes a vector of speaker representation inferred by a pretrained speaker encoder network to synthesize target speaker's speech. Importantly, the Glow model establishes an invertible mapping between the semantic representations and the speaker representations, facilitating speaker generation from text prompt. The process of language-guided speaker generation starts with predicting the prior semantic distribution based on the input text description and sampling to obtain the semantic representation. Then the Glow model transforms the semantic representation into a speaker representation, and finally the zero-shot TTS system synthesizes the generated speaker's speech based on the generated speaker representation. In this way, PromptSpeaker can generate new speakers according to user's natural language description on speaker characteristics, including gender, age and fine-grained speaker attributes.




PromptTTS~\cite{ref_prompttts} is the most related work that takes text descriptions to guide the speech synthesis process. The approach directly feeds the encoded text representations into the TTS system. Thus it is not suitable for the speaker generation task because it ignores the one-to-many mapping between the text prompt and the speaker timbre. In other words, the prompt ``a man's voice" can correspond to all male speakers in the training data, making it difficult for the model to establish a one-to-one correspondence between the prompt and the specific speaker timbre. The prompt can only represent some typical characteristics of the speaker's timbre, such as gentle, and low speaking speed, but cannot define the speaker's fine-grained timbre. As mentioned, since there is a one-to-many mapping between prompt and speaker timbre, in PromptSpeaker, we synthesize the distribution of a speaker representation based on the prompt. We sample this prior distribution to obtain the speaker representations not present in the training set. However, speaker representations do not match the semantic scale, i.e., two speakers' representations are not necessarily close even if they have similar timbre characteristics. Therefore, we use the Glow~\cite{ref_glow} model to establish an invertible mapping between speaker representations and semantic representations, realizing new speaker generation based on text descriptions.


\vspace{-3pt}
\section{Method}
\label{sec:Method}
\vspace{-3pt}

As shown in Fig.~\ref{model}, PromptSpeaker consists of a prompt encoder, a zero-shot TTS system and a Glow model. The prompt encoder produces the mean and variance for the semantic prior distribution. The zero-shot TTS system aims to take the speaker representation as input to synthesize the target speaker's speech. The Glow model establishes an invertible transformation between semantic and speaker features to achieve cross-modal mapping. To generate a new speaker, we first sample from the semantic prior distribution and transform it to speaker representations by the Glow model. Subsequently, the zero-shot TTS system takes the speaker representation as input to synthesize speech of the generated speaker. We will describe these modules in detail.


\subsection{Prompt Encoder}
\label{sec2:Prompt Encoder}
 \vspace{-3pt}
The prompt encoder extracts semantic information from speaker prompt with BERT~\cite{ref_bert} to predict the distribution of semantic prior representations. As shown in Fig.\ref{model}~(c), the prompt encoder consists of a pre-trained BERT model, FFT blocks~\cite{ref_fastspeech}, a GRU layer, a token layer~\cite{ref_gst} and a linear layer. The text-based speaker prompt is first fed into the BERT model to extract semantic features. Then, we feed the semantic features into a GRU layer to compress the sequence features into a vector to obtain a global-level speaker timbre representation. Subsequently a token layer is followed to further simplify the semantic features. The last linear layer finally produces the mean and variance of the semantic prior distribution.  


\subsection{Zero-Shot TTS System}
\label{sec2:zero-shot TTS system}

 \vspace{-5pt}
Since we need to synthesize speech based on the speaker representation of a generated \textit{unseen} speaker, a zero-shot TTS system is desired. Due to the success of the VITS architecture~\cite{ref_VITS} in the zero-shot task~\cite{ref_yourtts}, we take VITS with a pre-trained speaker encoder~\footnote[2]{\url{https://github.com/resemble-ai/Resemblyzer}} as our zero-shot TTS system. 

VITS adopts a conditional variational autoencoder (CVAE) structure, which includes three parts: a posterior encoder, a decoder, and a conditional prior network. The posterior encoder encodes the linear spectrogram to latent variables, and the decoder generates waveform based on the latent variables. The conditional prior network constrains the latent variables' encoding process and establishes the mapping of text to latent variables. In the speech synthesis process, the input text is mapped to latent features by the conditional prior network and then upsampled to the waveform. Specifically, VITS uses a flow model to improve the flexibility of the prior distribution.

Similar to YourTTS~\cite{ref_yourtts}, we adopt the speaker encoding method~\cite{ref_voiceclone} to construct the zero-shot VITS model. A well-trained zero-shot VITS can synthesize speech based on speaker representation of an unseen speaker. In the speaker generation process, the zero-shot VITS accepts a generated speaker representation to synthesize the speech of a speaker that does not exist in the training set.

\subsection{Glow}
\label{sec2:Glow}
 \vspace{-5pt}

We use a speaker encoder pre-trained on the speaker classification task to extract speaker representations, but the distribution of speaker representations does not necessarily match the distribution of semantic features, i.e., speakers satisfying a certain feature are not necessarily similar in speaker representations. In order to establish a matching association between speaker representations and semantic representations, we introduce the Glow model to PromptSpeaker. In the training process, the Glow model transforms the speaker representation into a semantic representation that matches the prompt prior distribution. In the inference process, the semantic representations sampled from the prompt prior distribution is transformed by the inverse of Glow to obtain the speaker representation.


\begin{table*}[h]
\centering
\caption{Dataset with text description on speaker voice characteristics}
\begin{tabular}{cccc}
\hline
dataset            & speakers & \begin{tabular}[c]{@{}c@{}}samples\\ per speaker\end{tabular} & \begin{tabular}[c]{@{}c@{}}text description\\ example\end{tabular} \\ \hline
Internal Stylistic & 74       & 100                                                           & \textit{a husky voice from a middle-aged man}                   \\
AISHELL-3~\cite{ref_aishell3}          & 218      & 20                                                            & \textit{voice from a middle-aged man}                                       \\
DiDiSpeech~\cite{ref_didispeech}         & 500      & 20                                                            & \textit{a boy's voice}                                           \\ \hline
\end{tabular}
\label{dataset}
\vspace{-5pt}
\end{table*}


\subsection{Training and Inference}
\label{sec2:training and inference process}

During the training process, since there is a one-to-many mapping between text description and speaker timbre, we use the speaker representations extracted by the speaker encoder as the input of the TTS system. In this way, we can ensure that the input of the TTS system contains complete speaker information. In the inference process, we obtain the prior distribution based on the text description of a desired speaker and sample from the prior distribution. The Glow model transforms the semantic representation into the speaker representation. The zero-shot TTS system synthesizes the generated speaker's speech based on the input text and the generated speaker representation.


\section{Dataset}
\label{sec:Dataset}

We built a multi-speaker Mandarin dataset with natural language descriptions of speaker characteristics. The speaker generation task requires diverse data from enormous speakers, so our dataset is composed of internal and open-source data simultaneously, as summarized in Table~\ref{dataset}. Specifically, we first manually annotate the high-quality internal data from 74 stylistic speakers, as described in Table~\ref{dataset}. Different people may have different descriptions of the same speaker, so each speaker is annotated by 13 annotators, reflecting the one-to-many mapping between text description and speaker timbre. Additionally, we use two open-source datasets, AISHELL-3~\cite{ref_aishell3} and DiDiSpeech~\cite{ref_didispeech}, to increase the number of speakers. Both datasets provide gender and age annotations for each speaker, and we automatically construct simple text descriptions based on these annotations, such as ``\textit{voice from a middle-aged man}", ``\textit{a boy's voice}", etc. To balance the data coverage, we do not use all the audio from the open-source data but only 20 sentences per speaker.

\vspace{-5pt}
\section{Experiments}
\label{sec:experiment}

\subsection{Model Configuration}
\label{sec3:Model Configuration}

We train the baseline and proposed models with the following model configuration for comparison.

\begin{itemize}
  \item
  \textbf{Baseline}: The baseline model consists of a prompt encoder and VITS. Similar to the structure of PromptTTS~\cite{ref_prompttts}, the baseline model feeds the encoded semantic features directly into the VITS system as speaker representations, and no speaker ID or extracted speaker embedding are used. The baseline model takes the same prompt encoder structure as the proposed model. During the inference process, the VITS model synthesizes speech based on the input text and the extracted semantic representations. Since the baseline model does not have the ability to obtain multiple speakers by sampling, we compare with the baseline model only in terms of speaker timbre accuracy when given a speaker prompt.
  
  \item
  \textbf{PromptSpeaker}: The proposed speaker generation system. The prompt encoder uses a pre-trained BERT model without finetune to extract word embedding and feed it into two layers of FFT blocks. The hidden dimension and filter dimension of FFT blocks are 256 and 1024, respectively. The hidden dimension of GRU is 256, and the token layer has 10 tokens. The dimensions of both semantic representation and speaker representation are 256. The Glow model contains 12 flow blocks and each flow block contains an activation normalization layer, an invertible $1\times1$ convolution layer, and an affine coupling layer. The kernel size and layer number of the affine coupling layer are 1 and 4, respectively.

\end{itemize}

The baseline and proposed models are trained up to 500K steps with a batch size of 12. To improve the performance of the zero-shot TTS system, we use a pre-trained VITS model as the initial model for the zero-shot VITS model. The pre-trained VITS model is trained using 500 hours of data from 250 speakers.



\subsection{Evaluation Metric}
\label{sec3:Evaluation Metric}

We evaluate the two models in terms of subjective and objective metrics. The subjective metrics consist of the \textit{naturalness MOS} of the synthesized audio and \textit{speaker timbre matching MOS}. Specifically, we ask 10 listeners to evaluate whether the generated speaker's timbre in synthetic speech matches the provided 20 test text descriptions, with 0 representing a total mismatch and 5 representing a perfect match. In addition to this, we also count the accuracy of the generated speaker's gender matching the text description. 

We also compute objective metrics to demonstrate whether the generated speaker differs from the speakers in the training set. As there is no typical standard on how difference between the two speakers, we follow the method in~\cite{ref_speakergeneration}, which designs several metrics and measures the difference between speakers based on the \textit{d-vector}. Specifically, we take WeSpeaker~\cite{ref_wespeaker}, a powerful speaker embedding model, to extract the d-vector from speech utterance and average the d-vectors extracted over ten sentences to obtain speaker-level d-vectors. The cosine similarity between speaker-level d-vectors is used as a distance metric between two speakers. We compute speaker-level d-vectors~$V$ for three types of speakers:

\begin{itemize}
\item
    \textbf{Ground truth training speaker~(gt)}: we compute $V^{gt}_{i}$ for each training speaker~$i$ on the ground-truth audio.
    
\vspace{1pt}
\item
    \textbf{Synthesized training speaker~(syn)}: we compute $V^{syn}_{i}$ for each training speaker~$i$ on the synthesized audio.

\vspace{1pt}
\item
    \textbf{Generated speaker~(gen)}: we compute $V^{gen}_{i}$ for the generated speaker~$i$ when given the test speaker prompt.

\end{itemize}



\begin{table*}[h]
\centering
\caption{Objective metrics on the quality of speaker generation}
\begin{tabular}{cclcccc}
\hline
\multicolumn{2}{c}{speaker fidelity} &  & \multicolumn{4}{c}{speaker generation}                   \\ \cline{1-2} \cline{4-7} 
syn2gt-same       & syn2gt-near      &  & syn2syn-same & syn2syn-near & gen2syn-near & gen2gen-near \\ \hline
0.143             & 0.252            &  & 0.024        & 0.113        & 0.085       & 0.088   \\ \hline
\end{tabular}
\label{tabel:objective}
 \vspace{-5pt}
\end{table*}

We average the speaker embedding extracted from the ten sentences to obtain the speaker-level d-vector and calculate the cosine distance $d(V_{1}, V_{2})=1-\frac{V_{1}}{\Vert V_{1}\Vert} \cdot \frac{V_{2}}{\Vert V_{2}\Vert}$ as the distance~\cite{ref_cosdis} between two speakers. When the cosine distance~$d(V_{1}, V_{2})$ between two speaker-level d-vectors is larger than a threshold, we consider $V_{1}$ and $V_{2}$ to be different speakers. Similar to the previous speaker generation work~\cite{ref_speakergeneration}, we define this threshold by computing the cosine distance between the training speakers when they are synthesized. We define the set of training speakers as $T$ and the set of generated speakers as $G$, and the following metrics are calculated:

\begin{itemize}
\item
    \textbf{syn2syn-same}: the distance between the same training speaker.
 \vspace{-5pt}
    \begin{equation}
    \underset{i \in T}{mean}~d(V_{i}^{syn}, V_{i}^{syn})
    \end{equation}
 \vspace{-10pt}
\item
    \textbf{syn2syn-near}: the distance between the closest  different synthesized training speakers.
 \vspace{-5pt}
    \begin{equation}
    \underset{i\in T}{mean}~\underset{j\in T,i\neq j}{min}~d(V_{i}^{syn}, V_{j}^{syn})
    \end{equation}
 \vspace{-10pt}
\item
    \textbf{gen2syn-near}: the distance between the generated speaker and the closest synthesized speaker.
 \vspace{-5pt}
    \begin{equation}
    \underset{i\in G}{mean}~\underset{j\in T}{min}~d(V_{i}^{gen}, V_{j}^{syn})
    \end{equation}
 \vspace{-10pt}
\item
    \textbf{gen2gen-near}: the distance between the closest generated speakers when given the same speaker prompt.
 \vspace{-5pt}
    \begin{equation}
    \underset{i\in G}{mean}~\underset{j\in G,i\neq j}{min}~d(V_{i}^{gen}, V_{j}^{gen})
    \end{equation}
 \vspace{-10pt}
\end{itemize}

\noindent The \textit{same} means we compute the metrics on the same speaker and the \textit{near} means that we compute the metrics on different speakers and select the one with the smallest distance. Different from Speaker Generation~\cite{ref_speakergeneration}, we additionally computed the distance between the same synthesized training speaker \textit{syn2syn-same} to specify the threshold of speaker distance. The \textit{gen2syn-near} represents the distance between the generated speaker and the nearby training speaker. When \textit{gen2syn-near} is close to \textit{syn2syn-same}, we assume that the generated speaker is one of the training speakers, i.e., the model does not generate a new speaker. When \textit{gen2syn-near} is close to \textit{syn2syn-near}, we consider that the generated speaker is not one of the training speakers, i.e., a new speaker is generated.

Following~\cite{ref_speakergeneration}, we also evaluate speaker fidelity by computing the distance between the synthesized and ground truth training speakers. We compute:

\begin{itemize}
\item
    \textbf{syn2gt-same}: the distance between the synthesized training speaker and the same ground truth training speaker.
 \vspace{-5pt}
    \begin{equation}
    \underset{i\in T}{mean}~d(V_{i}^{syn}, V_{j}^{gt})
    \end{equation}
 \vspace{-10pt}
\item
    \textbf{syn2gt-near}: the distance between the synthesized training speaker and the closest ground truth training speaker.
 \vspace{-5pt}
    \begin{equation}
    \underset{i\in T}{mean}~\underset{j\in T,i\neq j}{min}~d(V_{i}^{syn}, V_{j}^{gt})
    \end{equation}
 \vspace{-10pt}
\end{itemize}




\noindent When syn2gt-same is smaller than syn2gt-near, we believe that the zero-shot system can synthesize the corresponding speaker's timbre based on speaker embedding.


\subsection{Experimental Results}
\label{sec3:Experimental}

\subsubsection{Objective evaluation results}
\label{sec3:Objective evaluation results}

To evaluate whether the generated speakers appear in the training set, we calculate the speaker distance metrics listed in Section~\ref{sec3:Evaluation Metric}. We use 20 test text descriptions to generate speakers. As shown in Table~\ref{tabel:objective}, we can see that \textit{syn2gt-same} is smaller than \textit{syn2gt-near}, indicating that the zero-shot system can synthesize the corresponding speaker's speech based on speaker embedding. We can also see that gen2syn-near is between \textit{syn2syn-same} and \textit{syn2syn-near} and is closer to \textit{syn2syn-near}, so we believe that the generated speaker is more likely to be a speaker that is not in the training set. We also compute the distance \textit{gen2gen-near} between nearby speakers when we generate multiple speakers through sampling for a given text description, and we can see that \textit{gen2gen-near} is larger than \textit{syn2syn-same}, so we consider that the generated speaker timbres are diverse. In addition, \textit{syn2gt-same} is larger than \textit{syn2syn-same}, which indicates that there is still a difference between the speech synthesized by the TTS system and the real speech.

\begin{table}[]\centering
\caption{The naturalness MOS accessed on ground truth speech and synthetic speech from the training and generated speakers}
\begin{tabular}{cccc}
\hline
 & \begin{tabular}[c]{@{}c@{}}ground\\ truth\end{tabular} & \begin{tabular}[c]{@{}c@{}}training\\ speaker\end{tabular} & \begin{tabular}[c]{@{}c@{}}generated\\ speaker\end{tabular} \\ \hline
MOS   & 4.25~$\pm$~0.10  & 3.68~$\pm$~0.10 & 3.53~$\pm$~0.09  \\ \hline
\end{tabular}
\label{tabel:mos}
\end{table}

 
\subsubsection{Subjective  evaluation results}
\label{sec3:Subjective  evaluation results}
In a subjective evaluation, we assess the quality of the generated audio and the accuracy of the generated speaker when given a textual description. We use 20 test text descriptions to generate speakers, and one speech sample is generated for each text description. As shown in Table~\ref{tabel:mos}, we can see that there is no significant decrease in the naturalness MOS of the generated speakers compared to the training speakers. But there is still a gap between the synthetic speech and the ground truth speech. This is largely attributed to the use of low-quality open-source data for training speaker expansion.

We further count the gender accuracy and speaking timbre matching MOS of the generated speakers for the given text descriptions. As shown in Table~\ref{tabel:desmos}, we can see that PromptSpeaker performs significantly better than baseline in gender accuracy and speaker timbre MOS. This also proves that there is a one-to-many mapping between text description and speaker timbre, and it is reasonable to establish a mapping between speaker representation and text prior distribution for speaker generation. We notice that there is still a gap between ground-truth and generated speaker in speaker timbre matching MOS. In other word, the generalization ability of the learned semantic-speaker-timbre space is limited due to limited training data on speaker text-prompt pairs.

\begin{table}[]\centering
\caption{Subjective metrics in terms of gender accuracy and speaker timbre matching MOS}
\begin{tabular}{ccc}
\hline
             & gender accuracy & speaker MOS \\ \hline
ground-truth & 100\%           & 4.34~$\pm$~0.15    \\
baseline     & 46\%            & 1.87~$\pm$~0.28   \\
proposed     & 96\%            & 3.23~$\pm$~0.20  \\ \hline
\end{tabular}
\label{tabel:desmos}
\end{table}




\section{Conclusions}
\label{sec:Conclusions}

In this work, we propose PromptSpeaker, a controllable speaker generation model based on text descriptions, to make the speaker generation process more user-friendly. PromptSpeaker encodes text descriptions as semantic representations and transforms the semantic representations into a novel speaker representation as input for the zero-shot TTS system to generate speech. Experiments show that PromptSpeaker can generate speech from speakers that do not exist in the training set, with reasonable naturalness and matching quality between language prompt and generated speaker timbre.
But mainly due to the data limitation, there is substantial room for improvement and the gap can be further mitigated by labeling a large dataset. Moreover, we will continuously improve the accuracy of the generated speakers by modeling fine-grained speaker attributes separately.


\clearpage

\bibliographystyle{IEEEbib}
\bibliography{strings,refs}

\end{document}